\documentclass[aps,pre,twocolumn,10pt,superscriptaddress,nofootinbib,balancelastpage]{revtex4-2}
\usepackage[latin1]{inputenc}
\usepackage{amsmath,amssymb,amsthm}
\usepackage{mathrsfs}
\usepackage[capitalise]{cleveref}
\usepackage{siunitx}
\usepackage{braket}
\usepackage{calc}
\usepackage{tabularx}
\usepackage{dsfont}
\usepackage{color}
\usepackage{ifthen}
\usepackage[normalem]{ulem}% Fuer Durchstreichungen
\usepackage{graphicx}
\usepackage{listings}
\usepackage{float}

\newfloat{lstfloat}{htbp}{lop}
\floatname{lstfloat}{Listing}

\allowdisplaybreaks

\newcommand{\dif}{\mathrm{d}}%
\newcommand{\ZT}[1]{\textquotedblleft#1\textquotedblright}%

\newcolumntype{Y}{>{\centering\arraybackslash}X}%
\newcolumntype{Z}{>{\raggedright\arraybackslash}X}%

\newlength{\myl}%
\newcommand{\SUM}[2]{{\setlength{\myl}{\widthof{$\displaystyle\sum_{#1}^{#2}$}*\real{0.5}-\widthof{$\displaystyle\sum$}*\real{0.5}}\sum_{#1}^{#2}\;\hspace{-\the\myl}}}% Summen in abgesetzten Gleichungen
\newcommand{\INT}[3]{\settowidth{\myl}{$\displaystyle\int_{#1}^{#2}$}{\int_{#1}^{#2}\;\;\;\hspace{-\the\myl}\dif #3}\,}% Integrale in abgesetzten Gleichungen
\newcommand{\TINT}[3]{\settowidth{\myl}{$\int_{#1}^{#2}$}{\int_{#1}^{#2}\!\ifthenelse{\equal{#1#2}{}}{}{\;\;\;\;\hspace{-\the\myl}}\dif #3}\,}%
\newcommand{\EINT}[3]{\settowidth{\myl}{$\int_{#1}^{#2}$}{\int_{#1}^{#2}\;\;\;\,\hspace{-\the\myl}\dif #3}\,}% Integrale in Exponenten

\newtheorem{definition}{Definition}

\begin{document}

\title{Generic framework for data-race-free many-particle simulations on shared memory hardware}

\author{Julian Jeggle}
\affiliation{Institut f\"ur Theoretische Physik, Center for Soft Nanoscience, Westf\"alische Wilhelms-Universit\"at M\"unster, 48149 M\"unster, Germany}

\author{Raphael Wittkowski}
\email[Corresponding author: ]{raphael.wittkowski@uni-muenster.de}
\affiliation{Institut f\"ur Theoretische Physik, Center for Soft Nanoscience, Westf\"alische Wilhelms-Universit\"at M\"unster, 48149 M\"unster, Germany}

\begin{abstract}
Recently, there has been much progress in the formulation and implementation of methods for generic many-particle simulations. These models, however, typically either do not utilize shared memory hardware or do not guarantee data-race freedom for arbitrary particle dynamics. Here, we present both a abstract formal model of particle dynamics and a corresponding domain-specific programming language that can guarantee data-race freedom. The design of both the model and the language are heavily inspired by the Rust programming language that enables data-race-free general-purpose parallel computation. We also present a method of preventing deadlocks within our model by a suitable graph representation of a particle simulation. Finally, we demonstrate the practicability of our model on a number of common numerical primitives from molecular dynamics.
\end{abstract}
\maketitle

\section{Introduction}\label{introduction}
From the movement of gravitationally coupled celestial objects \cite{Fischer2014} to swarming animals \cite{Reynolds1987} to bacteria \cite{Zhang2010} and colloids \cite{Denz2021} to molecules in a fluid \cite{Rahman1964}, there are countless examples of naturally occurring and synthetically produced many-particle systems, i.e.,\ systems of interacting discrete entities. While it is typically easy to describe the dynamics of single particles analytically, the dynamics of many interacting particles is often only accessible numerically. Therefore, simulations of many-particle systems have become a standard tool for scientific computation and have numerous applications in physics \cite{Kong2011,Ristow1992}, chemistry \cite{vanGunsteren1990,vanDuin2001}, biology \cite{Lee2006,Perez2012}, and mathematics \cite{Sakuma2017,Edvardsson2015}.

Despite the fact that these kinds of simulation methods have been in use since the dawn of scientific computer simulations \cite{Alder1959}, there is still active development in the field \cite{Schwantes2014, Hertig2016, Hollingsworth2018}. One reason for this is that many-particle simulations frequently involve very computationally intensive calculations and therefore require careful algorithmic optimization. The complexity of crafting efficient implementations of many-particle dynamics has led to the development of numerous software packages, which promise to alleviate some of this complexity. This is especially the case in the field of molecular dynamics which has brought forward software packages such as GROMACS \cite{Pall2014}, HOOMD-blue \cite{Anderson2020}, LAMMPS \cite{LAMMPS}, or NAMD \cite{Phillips2020}.

In recent times, there has been an increased interest in more general particle simulation software. This has given rise to software packages like FDPS \cite{Iwasawa2016}, PPM \cite{Awile2013}, or OpenFPM \cite{Incardona2019}. These software packages are not simply a library of common routines like many molecular dynamics packages, but rather a framework for implementing custom particle dynamics, thus giving the user much more control over the exact numerical details of the simulation. Most of these existing software solutions for many-particle simulations utilize the distributed memory model for parallelization, i.e.,\ the particle system is decomposed into multiple subsystems, which are then processed in parallel. While this means that the simulation can be spread evenly across multiple processing nodes, it also comes with an inherent communication overhead due to the need to send information between particle subsystems \cite{Kale1999}.

At the same time, we see a trend in computer hardware towards increasingly parallel computing devices, e.g.,\ in the form of processors with increasingly large numbers of cores or specialized hardware such as graphics processors. This makes it viable to utilize a shared memory model on a single processing node instead, where all threads have access to the same memory. Some modern molecular dynamics frameworks such as OpenMM \cite{Eastman2017}, thus target \textit{shared-memory} CPU and GPU architectures to exploit current hardware to its fullest extent. While the shared-memory model eliminates most of the communication overhead, it comes at the cost of a parallelism model that is much more prone to errors, in particular a class of errors called \textit{data races} \cite{Netzer1989}.

While data races are highly problematic in general computation, their impact can be greatly reduced by limiting computation to a specific problem domain. The most popular example of this phenomenon are graphics shaders, which by the nature of graphics processors run on shared memory hardware, but are rarely affected by data races. This is due to the design of the render pipeline that restricts access to data as it is being processed in parallel \cite{Segal2019}.  This approach to preventing data races is taken even further in the novel systems programming language Rust \cite{Klabnik2019}, which generalizes the idea of access restrictions on variables such that general-purpose computation is possible.

Inspired by the success of Rust, we present in this article a domain-specific model and programming language for many-particle simulations with the goal of data-race freedom on shared-memory hardware. We show that many common particle dynamics can be expressed within this model and can therefore be parallelized safely. Our work can be seen as a continuation of Ref.\ \cite{Bamme2021}, however, we found it more productive to construct our model independently instead of maintaining compatibility with the notation used in Ref.\ \cite{Bamme2021}. Due to a very broad definition of what constitutes a \textit{particle dynamics}, the resulting model for safely parallel molecular dynamics surpasses the flexibility of existing shared-memory implementations such as OpenMM \cite{Eastman2017}.

This article is structured as follows: In Sec.\ \ref{sec:data_races_and_rust}, we briefly review the concept of data races and strategies to prevent them. Then we formulate a model to ensure data-race-free particle dynamics in Sec.\ \ref{sec:data_race_free_simulations}, before concretizing this model in form of a domain-specific programming language in Sec.\ \ref{sec:DSL}. We conclude in Sec.\ \ref{sec:conclusions}.

\section{Data races and data-race freedom}\label{sec:data_races_and_rust}
In this section, we briefly introduce the problem of \textit{data races} in parallel programming and present a number of methods to ensure their absence in programs \cite{Rauber2013,Klabnik2019}.

Generally speaking, programs can be viewed as sets of \textit{tasks}.\footnote{Sometimes these tasks are also called \textit{instructions}, which, however, might be confused with \textit{machine instructions}, which have additional technical connotations in computer architectures.} In a \textit{sequential} program, these tasks are executed in a predefined sequence called a \textit{thread of execution}, with only one task being worked on at the same time. In contrast, \textit{concurrent} programs do not have just a single thread of execution, but are composed of multiple threads of execution that can at least in principle progress independently. Notably, this does \textit{not} imply that multiple threads of execution advance at the same time. The program could instead simply switch between threads of execution periodically and advance them one at a time. A concurrent program, where multiple threads of execution do in fact advance \textit{simultaneously}, is called a \textit{parallel} program. \cite{Klabnik2019}

Compared to sequential programming, creating concurrent programs requires more diligence as one needs to take into account every possibility of the different threads of execution interleaving accesses to shared resources (e.g., memory). If the outcome of executing a program depends on the order of these accesses, the program is said to exhibit a \textit{race condition} as two or more threads \textit{race} for first access to a resource. A particularly significant case of this is a situation where two or more threads access the same location in memory concurrently with at least one thread \textit{writing} to memory and at least one access being \textit{unsynchronized}, i.e., not being forced into sequential execution with the other accesses. This situation is referred to as a \textit{data race} \cite{Klabnik2019}. Most programming language models give no guarantees to program behavior if a data race occurs (\textit{undefined behavior}). This, combined with the erratic nature of race conditions in general, makes data races a class of programming errors that poses a significant threat to program correctness.

Consequently, data race detection and avoidance is an important field of research \cite{Serebryany2009, Netzer1991}. One solution to avoid data races is to simply not share any memory. This is often used in the \textit{message-passing} scheme of parallel programming, where threads cannot share data, but can send data between each other. However, message passing is less efficient compared to sharing memory as the data sent between threads typically needs to be copied first to form a contiguous message. This creates a communication overhead that can degrade performance if too much information needs to be communicated.

Here, we want to focus on the solution to the problem of data races utilized by the Rust Programming Language (henceforth simply called \ZT{Rust}). Rust is a modern systems programming language with a focus on reliability and efficiency. On the one hand, Rust offers low-level access to details of execution, e.g., memory management or operating system level multithreading, like typical systems programming languages such as C or C++. On the other hand, it also comes with strong correctness guarantees via a powerful type system and static analysis checks. In fact, in the so-called \textit{Safe Rust} subset of Rust, it is \textit{impossible} to trigger undefined behavior in any legal program.

This safety is possible in part due to the concepts of memory \textit{ownership} and reference \textit{lifetimes} being explicit and pervasive in Rust programs.
At compile time the Rust compiler validates programs with respect to a set of rules regarding these concepts that are designed specifically to prevent data races as well as other memory-related sources of undefined behavior. In brief, Rust prohibits \textit{aliasing} of values, i.e., at every point in time every value (i.e., some block of memory filled with data) of a Rust program is bound to exactly one variable. While the variable is in scope, the value must remain allocated, and conversely, when the variable drops out of scope, the value is freed. This prevents \textit{use-after-free} errors without the overhead of memory management via \textit{garbage collection} \cite{Barth1977}. Besides value ownership, Rust also allows \textit{borrowing} a value from a variable as a reference. To prevent use-after-free errors, the Rust compiler uses an elaborate system of tracking \textit{lifetimes} of references to prove that no reference cannot outlive the variable it is derived from.

Of particular importance for this work is secondary restriction imposed on references in Rust, the so-called \textit{rules of borrowing}. These state that at every point in time, there can be either an arbitrary number of read-only references or a single mutable reference. In either case, data races are impossible: there is either no writing access or only a single thread carries a reference to the same location in memory. It is this idea of regulating concurrent access and mutability of data that forms the basis for our domain-specific model for safely parallelizable particle simulations. One should keep in mind that all this reasoning has to happen at compile time as to not incur a management overhead that degrades performance.

It is outside the scope of this article to explain the intricacies of Rust in more detail, so we instead refer the interested reader to Refs. \cite{Jung2017,Weiss2019,Jung2021,Klabnik2019} for in-depth explanations on the design of Rust.

At this point it should be highlighted that Rust does not provide its safety guarantees for free. Programmers have to handle the increased complexity of designing their programs in such a fashion that the compiler can track ownership and reference lifetimes. Because this is a special case of static program analysis, the Rust compiler is doomed by the impossibility of \textit{perfect} static program analysis \cite{Rice1953} to always reject some programs even though they are in fact (provably) safe.\footnote{In fact, it is trivial to find such a program in Rust. Because Rust treats arrays as a single value with respect to the aliasing rules the Rust compiler will reject a multithreaded program even if each thread accesses only a mutually disjoint subset of the array.} To circumvent this, Rust allows programmers to selectively opt out of some static analysis checks via so-called \textit{Unsafe Rust}. This, of course, places the responsibility of validating safety on the programmer, but can be used very effectively to create safe abstractions over inherently unsafe operations. This ``escape hatch'' design of Rust is another aspect we take inspiration from in this article.

Before we discuss the specific semantics of simulating particle systems numerically we will first present a more general formalization of the idea of safe parallelism via thread-exclusive mutability of variables. For this purpose, we first define the notion of a \textit{program state} as the collective state of all variables of a program. To distinguish between different variables we identify each variable with a natural number. In the interest of simplicity we do not differentiate between different variable types but instead only assume all variables can take values from some generic set.

\begin{definition}
    A \textbf{program state} is a function $\sigma:I\rightarrow V$ that maps a finite index set $I\subset\mathbb{N}$ to the set of possible variable values $V$. The value of the $i$-th variable is represented as $\sigma(i)$. The set of all possible program states with a given index set $I$ and value set $V$ is $V^{I}$, i.e., the set of all functions that map from $I$ to $V$.
\end{definition}

It should be noted that unlike typical automata models such as register machines this definition of a program state lacks any notion of a instruction counter or a similar construct to track the flow of execution. Instead, we describe program flow extrinsically as a sequence of fixed program steps.

\begin{definition}
    A \textbf{program step} for a set of possible variable values $V$ is defined as a tuple $(I,i,\delta)$ with an index set $I\subset\mathbb{N}$, a variable index $i\in\mathbb{N}$ and an update function $\delta\in(V^{I}\rightarrow V)\cup\{\mathrm{DEL}\}$ which must satisfy the condition $\delta=\mathrm{DEL}\Rightarrow i\in I$. Conceptually, this encodes a change in a program state $\sigma$ with index set $I$ where the $i$-th variable is substituted by $\delta(\sigma)$. If $i\notin I$, the variable is newly created and initialized by $\delta(\sigma)$ and if $\delta=\mathrm{DEL}$ the variable is deleted instead. We define the output index set $I_{\mathrm{out}}\subset\mathbb{N}$ as 
    \begin{align*}
        I_{\mathrm{out}}((I,i,\delta))=\begin{cases}
            I & i\in I\wedge\delta\neq\mathrm{DEL}\\
            I\backslash\{i\} & i\in I\wedge\delta=\mathrm{DEL}\\
            I\cup\{i\} & i\notin I\wedge\delta\neq\mathrm{DEL}
            \end{cases}
    \end{align*}
    To map input program states to output program states we define the \textbf{execution function} $\mathrm{exec}((I,i,\delta)):V^{I}\rightarrow V^{I_{\mathrm{out}}((I,i,\delta))}$ as 
    \begin{align*}
        \mathrm{exec}((I,i,\delta))(\sigma)(j)=\begin{cases}
        \sigma(j) & i\neq j\\
        \delta(\sigma) & i=j\wedge\delta\neq\mathrm{DEL}
        \end{cases}
    \end{align*}
\end{definition}

\begin{definition}
    Let $p=(I_{1},i_{1},\delta_{1}),\dots,(I_{n},i_{n},\delta_{n})$ be a finite sequence of length $n\in\mathbb{N}$ such that $(I_{k},i_{k},\delta_{k})$ is a program step for all $1\leq k \leq n$. We call this sequence a \textbf{program} if the index set of each program step is the same as the output index set of the previous program step, i.e., for all $1\leq k<n$ it holds that $I_{k+1}=I_{\mathrm{out}}((I_{k},i_{k},\delta_{k}))$. We define the \textbf{execution function} of the program $\mathrm{exec}(p):V^{I_{1}}\rightarrow V^{I_{\mathrm{out}}((I_{n},i_{n},\delta_{n}))}$, i.e., the function that maps from the initial state to the output state after executing each program step in order, as the composition of the execution functions of each program step
    \begin{align*}
        \mathrm{exec}(p)=\mathrm{exec}((I_{n},i_{n},\delta_{n}))\circ\dots\circ\mathrm{exec}((I_{1},i_{1},\delta_{1})).
    \end{align*}
    Furthermore we define the \textbf{input index set} of the program as $I_{\mathrm{in}}(p)=I_{1}$ and the \textbf{output index set} of the program as $I_{\mathrm{out}}(p)=I_{\mathrm{out}}((I_{n},i_{n},\delta_{n}))$.
\end{definition}

This way of modeling programs comes with the obvious drawback that the program flow must be the same regardless of program input. While this is a severe limitation for general computing, it is much less problematic in the context of molecular dynamics, where the program flow is typically determined only by the underlying model equations and not the concrete state of a given physical system. It should also be noted that the static nature program flow in the above definition does not rule out branching computation in general since any kind of computation can be incorporated into the update functions $\delta$.\footnote{Naturally, for this model to be of any practical use we have to assume that $\delta$ is computable (if $\delta\neq\mathrm{DEL}$).} As we will see later, this will form an \ZT{escape hatch} in our domain-specific programming language similar to Unsafe Rust.

Using the definition of a program in terms of its access patterns to variables we can now formalize the concepts of a program depending on and mutating variables.

\begin{definition}
    \label{def:program_mut}
    Let $p=(I_{1},i_{1},\delta_{1}),\dots,(I_{n},i_{n},\delta_{n})$ be a program. We define the \textbf{set of variables mutated by} $p$ as $\mathrm{mut}(p)\subseteq I_{\mathrm{out}}(p)\cup I_{\mathrm{in}}(p)$ such that for all $i\in\mathrm{mut}(p)$ there must be a $1\leq k\leq n$ with $i=i_{k}$. In other words, a variable is mutable with respect to a program if there is at least one program step in this program that updates it.
\end{definition}

\begin{definition}
    Let $(I,i,\delta)$ be a program step. We say that this \textbf{step depends on a variable} $k$ if there are two program states $\sigma,\sigma'\in V^{I}$ with $\sigma(k)\neq\sigma'(k)$ and $\sigma(\ell)=\sigma'(\ell)$ for all $\ell\neq k$ such that $\delta(\sigma)\neq\delta(\sigma')$, i.e., there are two program states that differ only in the value for the $k$-th variable but produce different output states.
\end{definition}

We emphasize again that we do not make any assumption on how $\delta$ is computed. Therefore, a program step $(I,i,\delta)$ not depending on the $k$-th variable does not imply that \textit{every} implementation of $\delta$ will be free of reading operations for this variable, but merely implies the \textit{existence} of an implementation without such an operation. In practice, it is usually sufficient to analyze \textit{syntactically} if an implementation of $\delta$ reads a given variable, e.g., by finding the corresponding symbol for the variable in the program code.\footnote{This method of analysis allows for false positives as it does not verify the reachability of the expression containing the symbol in question. Therefore it does not violate the general undecidability of (perfect) static analysis.}

\begin{definition}
    \label{def:program_dep}
    Let $p=(I_{1},i_{1},\delta_{1}),\dots,(I_{n},i_{n},\delta_{n})$ be a program. For each variable $i\in I_{\mathrm{in}}(p)$ we can split the program into two programms $p_{\mathrm{ro},i}$ and $p_{\mathrm{rem},i}$ such that the following three properties hold:
    \begin{enumerate}
        \item $p_{\mathrm{ro},i}p_{\mathrm{rem},i}=p$
        \item $p_{\mathrm{ro},i}$ contains no element $(I_{k},i_{k},\delta_{k})$ with $i_{k}=i$
        \item $p_{\mathrm{ro},i}$ has maximum length
    \end{enumerate}
    In other words, $p_{\mathrm{ro},i}$ is the part of program $p$ before there is a write to the $i$-th variable and $p_{\mathrm{rem},i}$ contains the remaining program steps of $p$. We then define the \textbf{set of variables $p$ depends} on as $\mathrm{dep}(p)\subseteq I_{\mathrm{in}}(p)$ where $i\in\mathrm{dep}(p)$ if and only if $p_{\mathrm{ro},i}$ contains a step that depends on the variable $i$. Conceptually, $p$ depending on a variable means that during the execution of $p$ the value of the variable is read before it is first written to.
\end{definition}

Finally, we use Def.\ \ref{def:program_mut} and \ref{def:program_dep} to formally define the notion of data-race freedom by exclusive mutability as follows.

\begin{definition}
    \label{def:parallelizable_general}
    Let $p$ be a program $p=p_{1}\dots p_{n}$ composed of $n$ subprograms with $I_{\mathrm{in}}(p)=I_{\mathrm{in}}(p_{k})=I_{\mathrm{out}}(p_{k})=I_{\mathrm{out}}(p)=I$ for all $1\leq k\leq n$. We say that $p$ can be \textbf{parallelized without data races} via $p_{1},\dots,p_{n}$ if for all variables $i\in I$ the following conditions hold:
    \begin{enumerate}
        \item The variable $i$ is mutated by at most one subprogram, i.e., there is at most one $k\in\mathbb{N}^{\leq n}$ such that $i\in\mathrm{mut}(p_{k})$.
        \item If the variable is mutated by one subprogram, no other subprogram may depend on this variable, i.e., if there is a $k\in\mathbb{N}^{\leq n}$ such that $i\in\mathrm{mut}(p_{k})$ it is implied that for all $\ell\in\mathbb{N}^{\leq n}$ with $\ell\neq k$ it holds that $i\notin\mathrm{dep}(p_{\ell})$.
     \end{enumerate}
\end{definition}

The strategy we used to obtain Def.\ \ref{def:parallelizable_general} serves as a guideline for the next section, where we consider the problem domain of particle simulations.

\section{Data-race-free particle simulations}\label{sec:data_race_free_simulations}
In this section, we derive a general model for particle simulations for which we can make guarantees in terms of data-race freedom. To this end, we first define three concepts: \textit{physical quantities} associated with particles, \textit{particle types}, and \textit{particle systems}. With these we can then define \textit{particle dynamics} in a very general fashion and classify certain \textit{particle dynamics} as particularly useful.

\subsection{Modelling static particle systems}

Conceptually, a \textit{physical quantity} of a particle is a single, semantically atomic property of the particle, e.g., its position, mass, charge, velocity, or orientation. We make no assumptions on these quantities or their physical meaning aside from the fact that they can be represented by a finite number of real numbers.\footnote{For the sake of simplicity, we ignore the fact that physical quantities typically also possess a unit.}

\begin{definition}
    Let $n\in \mathbb{N}$. Then we call $Q=\mathbb{R}^n$ a \textbf{physical quantity type} of dimensionality $\textup{dim}(Q)=n$. $q\in Q$ is called a physical quantity of type $Q$. Furthermore, we call the \textbf{set of all physical quantity types} $\mathfrak{Q}=\{\mathbb{R}^n|n\in \mathbb{N}\}$ and the \textbf{set of all physical quantities} $\mathcal{I}(\mathfrak{Q})$, i.e., $q\in \mathcal{I}(\mathfrak{Q})$ implies the existence of an $n\in \mathbb{N}$ such that $q\in \mathbb{R}^n$.
\end{definition}

Generally, we can express the idea of a \textit{particle} as an entity that ties together a position in space with a varying number of physical quantities, e.g., orientation, (angular) velocity, mass, or charge. For a truly general model of particles we make no assumptions on the nature of these quantities except that their number must be finite.\footnote{Strictly speaking, this can be seen as a loss of generality. One could, e.g., imagine a particle that keeps a memory of its past states, e.g.\ to implement self-avoiding dynamics. Since a full representation of all past states of a particle cannot be expressed in a finite number of quantities, our model cannot capture these kinds of particles. However, practical implementations of particles with memory typically have either a temporal decay of memory (thus limiting the number of previous states that are remembered) or utilize global fields instead of per-particle memory to enable memory.} When defining this idea formally, we must distinguish between the concepts of \textit{particle types} and \textit{particle states}. The first can be seen as a blueprint for a set of particles equal in structure, while the second one encapsulates the idea of a single concrete particle at a given point in time.

\begin{definition}
    Let $I\subset\mathbb{N}$ be a finite set with $1\in I$ and let $\kappa:I\rightarrow\mathfrak{Q}$ be a function. Then $P=(I,\kappa)$ is a \textbf{particle type} with the index set $I$ and the quantity function $\kappa$. For a particle type $P=(I,\kappa)$ we call $\textup{pos}(P)=\kappa(1)$ the \textbf{position quantity type} of $P$ and $\textup{dim}(P)=\textup{dim}(\kappa(1))$ the \textbf{dimensionality} of the particle type. Furthermore, we define $\mathfrak{P}$ as the set of all particle types and $\mathfrak{P}_{n_\textup{dim}}$ as the set of all particle types of dimensionality $n_\textup{dim}$.
\end{definition}
\begin{definition}
    Let $P=(I,\kappa)$ be a particle type. Then we call $p:I\rightarrow\mathcal{I}(\mathfrak{Q})$ a \textbf{particle state} of type $P$ if for all $i\in I$ it holds that $p(i)\in \kappa(i)$, i.e., a particle state maps the index set $I$ to physical quantities in accordance to the quantity types defined in the corresponding particle type. For a particle state $p$ we define $\textup{pos}(p)=p(1)$ as the \textbf{position} of the particle state. Furthermore, we define $\mathcal{I}(P)$ as the set of all particle states of particle type $P$ as well as $\mathcal{I}(\mathfrak{P})$ as the set of all particle states of any particle type and $\mathcal{I}(\mathfrak{P}_{n_{\textup{dim}}})$ as the set of all particle states of any particle type of dimensionality $n_\textup{dim}$.
\end{definition}

The purpose of the index set $I$, which associates every quantity of a particle state with a unique index, may seem questionable to a reader at first glance. As we will see later, it is crucial, however, to reason about partial particle states, i.e., restrictions of a particle state onto a subset of quantities. $I$ can also serve to give physical meaning to quantities by some form of indexing scheme. This makes reasoning within the particle model much easier than, e.g., modelling particle states as (unordered) sets of quantities would do.

Finally, we can use the notion of particle types and states to define \textit{particle systems} and their states. Conceptually, a particle system contains two pieces of information. The first information contains the various particle types found in the system as well as how many particles of each type are contained within the system. Secondly, a particle system may also have physical information that is not bound to individual particles, i.e., information that is \textit{global} to the system. Examples of this are the simulation time or external fields. Again, we make no assumption on these global information except for the fact that they must be representable by a finite set of real numbers. One might ask at this point if the global state could not simply be represented as a single particle of a unique particle type. As we will see in the next section, separating the particle system from the global state is worthwhile to exploit the fact that for many numerical simulations of particle systems the global state is mutated separately from the state of the particle system.

\begin{definition}\label{def:particle_system}
    Let $I\subseteq\mathbb{N}$ be a finite set, $\tau:I\rightarrow\mathfrak{P}_{n_{\textup{dim}}}$ with $n_{\textup{dim}}\in\mathbb{N}$ and $\nu:I\rightarrow\mathbb{N}$ be functions as well as $G=\mathbb{R}^{m}$ for some $m\in\mathbb{N}_{0}$. Then $S=(I,\tau,\nu,G)$ is a \textbf{particle system} of dimensionality $n_{\textup{dim}}$ with the index set $I$, the particle type function $\tau$, the particle number function $\nu$, and the global-state space $G$. For each $i\in I$ we say that the system $S$ contains $\nu(i)$ particles of type $\tau(i)$.
\end{definition}
\begin{definition}
    Let $S=(I,\tau,\nu,G)$ be a particle system, $\sigma:I\rightarrow\mathcal{M}(\mathcal{I}(\mathfrak{P}))$ be a function mapping every element of $I$ to a multiset\footnote{See Appendix \ref{appendix_multisets} for a complete overview of the multiset formalism used here.} of particle states of any type and $g\in G$. Then we call $s=(\sigma,g)$ a \textbf{state of the particle system} $S$ if for all $i\in I$ and for all $p\in\sigma(i)$ it holds that $p\in\mathcal{I}(\tau(i))$ and $|\sigma(i)|=\nu(i)$. In other words, $\sigma$ is a function that maps the index for each particle type to a multiset containing the states for each particle of this type in the particle system. We define $\mathcal{I}(S)$ as the set of all possible states of a particle system $S$, i.e.\ $s\in\mathcal{I}(S)$ if and only if $s$ is a state of $S$.
\end{definition}

One should note that in the definitions above there is no concept of any ordering of particles, as the particle states for each particle type in the system are represented by a multiset. For the same reason, there is also no notion of particle identity inherent in the model, i.e., two or more particles in the same state are indistinguishable when represented in the state of a particle system. However, if desired, one can express both an ordering of particles and an inherent distinguishability of particles by extending the particle type with a (non-physical) quantity signifying particle identity (e.g., by using a quantity to enumerate particles).

To illustrate the above formalism, let us consider a particle system comprised of three point masses, i.e., particles that have position, velocity, and mass, in three spatial dimensions. Additionally, two of these particles shall also have an electric charge.\footnote{At this point, one could ask if it was not more sensible to unify all particles into one type, by expressing the uncharged particles as particles with zero charge. While sensible for this tiny example system, in larger systems, where the number of uncharged particles is high compared to that of the charged particles, it becomes computationally wasteful to use the more complicated dynamics of charged particles for the uncharged particles as well.} We say the particles have a position $\vec{r}_i$, a velocity $\vec{v}_i$, a mass $m_i$, and in the case of the charged particles a charge $q_i$ where $i=1$ represents the uncharged particle and $i\in\{2,3\}$ the charged particles. Both uncharged and charged particles can then be represented by these particle types, respectively,
\begin{align}
    P_{u}&=\left(\{1,2,3\},i\mapsto\begin{cases}
        \mathbb{R}^{3} & i\in\{1,2\}\\
        \mathbb{R} & i=3
        \end{cases}\right),\\
    P_{c}&=\left(\{1,2,3,4\},i\mapsto\begin{cases}
        \mathbb{R}^{3} & i\in\{1,2\}\\
        \mathbb{R} & i\in\{3,4\}
        \end{cases}\right),
\end{align}
and we can define the particle system as
\begin{align}
    S&=\left(\{1,2\},\tau,\nu,\{\}\right),\\
    \tau&:\{1,2\}\rightarrow\{P_{1},P_{2}\},i\mapsto\begin{cases}
        P_{u} & i=1\\
        P_{c} & i=2
        \end{cases},\\
    \nu&:\{1,2\}\rightarrow\mathbb{N},i\mapsto\begin{cases}
        1 & i=1\\
        2 & i=2
        \end{cases}.
\end{align}
Then we can express the state of the particle system as
\begin{align}
    s=(\sigma,g),\;\;\sigma(1)=[p_{1}],\;\;\sigma(2)=[p_{2},p_{3}]
\end{align}
with the particle states defined by
\begin{alignat}{3}
    p_{1}(1)&=\vec{r}_{1},& p_{1}(2)&=\vec{v}_{1}, & p_{1}(3)&=m_{1},\\
    p_{2,3}(1)&=\vec{r}_{2,3},\;\;& p_{2,3}(2)&=\vec{v}_{2,3},\;\;&p_{2,3}(3)&=m_{2,3},\\
    p_{2,3}(4)&=q_{2,3}.
\end{alignat}

\subsection{Modelling particle dynamics}\label{subsec:particle_dynamics}

Using the definitions from the previous section, we can define \textit{particle dynamics} simply as transitions between states of particle systems.

\begin{definition}\label{def:particle_dynamics}
    Let $S$ and $S'$ be particle systems. Then a function $d:\mathcal{I}(S)\rightarrow\mathcal{I}(S')$ is called a \textbf{particle dynamics}. For $s\in S$ we call $s'=d(s)$ the \textbf{evolved state} of $s$ under $d$.
\end{definition}

It should be noted that we do not impose \textit{any} restrictions on particle dynamics other than the fact that it maps from all valid states of \textit{some} particle system to valid states of \textit{some other} particle system. In particular, we do not assume that the particle system being mapped from is the same as the one being mapped to. This allows a particle dynamics to, e.g., change the number of particles in the system, alter particle types, and redefine global state.

An important algebraic property of particle dynamics as defined above is that they are closed under composition under the condition that input and output states are compatible. This is relevant since in practice particle simulations are often formulated as a loop over multiple operations each of which is considered elementary within some numerical scheme (e.g., applying a force all particles of a type or advancing particle positions). Our model can express this by defining a particle dynamics for each of these elementary operation and then compose them into more complex dynamics. As we will see later, this is highly useful as these elementary operations can typically be reasoned about more strongly than their composition into a complex particle dynamics.

On its own, the above definition of particle dynamics is far too generic to be of practical use. In a sense it is simply a (convoluted) way of expressing functions mapping between vector spaces of finite dimension. For this we cannot make any general statements regarding parallelism. Similar to how Safe Rust carves out of all possible programs only those programs for which it can guarantee safety, we can carve out of all possible particle dynamics only those which we can safely process in parallel in some fashion.

The first useful classification of particle dynamics distinguishes between dynamics that conserve global state and those that do not.
\begin{definition}
    Let $d:\mathcal{I}(S)\rightarrow\mathcal{I}(S')$ be a particle dynamics with $S=(I,\tau,\nu,G)$ and $S'=(I',\tau',\nu',G')$ being particle systems. $d$ is a \textbf{global-state preserving} particle dynamics if $G=G'$ and for all $(\sigma,g)\in \mathcal{I}(S)$ it holds that $d((\sigma,g))=(\sigma',g)$. In other words, the global state is invariant under the dynamics.
\end{definition}
This separation is useful as the time evolution of global state can be driven by \textit{any} numerical procedure. For example, the global state might contain discretized fields that are integrated in time by a numerical scheme such as finite differences. Parallel implementations of these generic algorithms are not specific to many-particle simulations and thus outside the scope of our simulation model for particle dynamics. Instead, it is more sensible to delegate the analysis and implementation of these kinds of programs to general purpose programming languages such as Rust.

Another classification of particle dynamics can be made by looking at the definition of particle types in the two particle systems connected by a particle dynamics. From Def.\ \ref{def:particle_dynamics} there is no requirement for both particle systems to contain the same particle types. In some special instances this can be useful, e.g., for initializing a system subject to a complex dynamics via a simple dynamics. For instance, one can use a simple diffusion dynamics to relax a system of particles into a homogeneous state to then run a more complex particle dynamics (with different particle types) from there. However, like in the case for dynamics not preserving global state, dynamics that do not preserve particle types are not algorithmically restricted enough to make statements about safe parallelism. We therefore characterize \textit{particle-type preserving dynamics} as particle dynamics that do not alter which particle types are found in a given particle system.
\begin{definition}
    Let $d:\mathcal{I}(S)\rightarrow\mathcal{I}(S')$ be a particle dynamics with $S=(I,\tau,\nu,G)$ and $S'=(I',\tau',\nu',G')$ being particle systems. $d$ is a \textbf{particle-type preserving} particle dynamics if $I=I'$ and $\tau=\tau'$.
\end{definition}

If we look at particle dynamics that are both global-state preserving and particle-type preserving, we can identify two subclasses of interest.

We consider first a class of particle dynamics that add particles of existing types to the particle system, but leaves existing particles untouched. To add particles to the system, we require the number of new particles as well as an initialization function for each particle type. Of particular interest for parallelization are dynamics of this kind, where the initialization of each new particle can be performed in parallel, i.e., independently from one another.
\begin{definition}\label{def:indep_init}
    Let $d:\mathcal{I}(S)\rightarrow\mathcal{I}(S')$ be a global-state preserving and particle-type preserving particle dynamics with $S=(I,\tau,\nu,G)$ and $S'=(I,\tau,\nu',G)$ being particle systems. Also, let $\zeta:I\rightarrow\mathbb{N}_{0}$ and $\eta:I\rightarrow\bigcup_{i\in I}(\mathbb{N}_{0}^{\leq\zeta(i)}\times\mathcal{I}(S)\rightarrow\mathcal{I}(\tau(i)))$ be functions indicating the number of new particles and their initial state for each particle type, respectively. We then call $d$ an \textbf{independently initializing insertion} if $\nu'(i)=\nu(i)+\zeta(i)$ and $d(\sigma,g)=(i\mapsto\sigma(i)+\sum_{k=1}^{\zeta(i)}[(\eta(i))(k,(\sigma,g))],g)$.
\end{definition}
Definition \ref{def:indep_init} formalizes the idea of an independent initialization of new particles by generating a dynamics from two functions $\zeta$ and $\eta$ where $\zeta$ determines the number of particles to create for each particle type and $\eta$ initializes a new particle for a given type based on an index enumerating each new particle as well as the state of the particle system before adding any particles. Since $\eta$ only depends on the original system state and the number of particles to create for each type is known in advance, all evaluations of $\eta$ can be performed in parallel without a data race as long as the new particles are only added to the system state after all particles have been initialized.

Next, we consider a class of particle deleting dynamics based on a predicate function that selects the particles to delete. Again, we can parallelize this dynamics safely if the individual evaluations of this predicate only depend on the individual particle state as well as the original system state.
\begin{definition}\label{def:indep_sel_del}
    Let $d:\mathcal{I}(S)\rightarrow\mathcal{I}(S')$ be a global-state preserving and particle-type preserving particle dynamics with $S=(I,\tau,\nu,G)$ and $S'=(I,\tau,\nu',G)$ being particle systems and $s=(\sigma,g)$ being a state of $S$. Also, let $\delta:I\rightarrow\bigcup_{i\in I}(\mathcal{I}(\tau(i))\times\mathcal{I}(S)\rightarrow\mathbb{B})$ be a Boolean predicate which flags particles for deletion. We then call $d$ an \textbf{independently selecting deletion} if $\nu'(i)=\nu(i)-|\textup{select}(\sigma(i),p\mapsto(\delta(i))(p,s))|$ and $d((\sigma,g))=(i\mapsto\textup{select}(\sigma(i),p\mapsto\neg(\delta(i))(p,(\sigma,g))),g)$.
\end{definition}
It should be noted that while for an independently initializing insertion the evolved particle system only depends on the original particle system, for an independently selecting deletion the evolved particle system depends both on the original particle system \textit{and} its state due to the fact that the number of deleted particles might depend on this state. Therefore, an independently selecting deletion as defined in Def.\ \ref{def:indep_sel_del} can only be meaningfully applied to a single state even if it is technically defined on all states of the original particle system. This is largely irrelevant for the domain-specific language developed later as we will design this language to be generic over particle numbers.

Until now, we have considered particle dynamics that can change the number of particles. Let us now look at particle-number conserving dynamics.
\begin{definition}
    Let $d:\mathcal{I}(S)\rightarrow\mathcal{I}(S')$ be a particle-type preserving particle dynamics with $S=(I,\tau,\nu,G)$ and $S'=(I,\tau,\nu',G')$ being particle systems. $d$ is a \textbf{particle-number preserving} particle dynamics if $\nu=\nu'$.
\end{definition}
By itself this class does not directly provide a parallelization opportunity as we have not specified how $d$ calculates the evolved state. Ideally, one would like to process every particle in parallel as this will provide ample opportunity for parallelization for most real-world systems. To this end we define a new subclass of particle dynamics.
\begin{definition}\label{def:particle_parallel}
    Let $d:\mathcal{I}(S)\rightarrow\mathcal{I}(S)$ be a global-state, particle-type and particle-number preserving particle dynamics with $S=(I,\tau,\nu,G)$ being a particle system. We call $d$ a \textbf{particle-parallel} particle dynamics under $\Pi$ if there is a function $\Pi:I\rightarrow\bigcup_{i\in I}(\mathcal{I}(\tau(i))\times\mathcal{I}(S)\rightarrow\mathcal{I}(\tau(i)))$ such that $d((\sigma,g))=(i\mapsto\textup{map}(\sigma(i),p\mapsto(\Pi(i))(p,(\sigma,g))),g)$. In other words, for every particle type $\tau(i)$ the function $\Pi$ produces a function $\Pi(i)$ that takes the state of a single particle of type $\tau(i)$ as well as the initial state of the particle system and produces the new state for this particle.
\end{definition}
It is noteworthy that, as a consequence of enforcing this form of dynamics, particles of the same type in the same state are necessarily mapped to the same state in the evolved system. We also enforce particles of the same type to be subject to the same state-mapping function, i.e., the notion of a particle type is now not only used to unify particles with a common set of quantities associated with them, but also particles that follow the same kind of behavior within a particle dynamics. While formally a restriction, this form of dynamics does not diminish the expressiveness of the particle model in practice. In particular, dynamics with completely unique behavior for each particle can be implemented by using a unique particle type for each particle. Similarly, if the restriction for particles of the same type in the same state is undesired, one can simply introduce a new quantity in the particle type to make particles of this type distinguishable.

A particle-parallel particle dynamics can be trivially parallelized by evaluating the respective $\Pi(i)$ for each particle in parallel. However, $\Pi(i)$ also has a secondary dependency on the state of the entire particle system. This is necessary to allow the dynamics of a single particle to depend not only on its own original state but also on the state of other particles, i.e., this enables particle \textit{interactions}. If we wish to mutate the system state in-place (i.e., without making a copy first)\footnote{Making a full copy of the entire particle system in each step of the simulation would be prohibitively expensive in terms of memory and runtime in many applications.} we might cause data races if we naively allow \textit{arbitrary} functions $\Pi(i)$ to run in parallel. To formulate a restriction on $\Pi(i)$ such that data-race freedom is guaranteed for in-place state updates, we first introduce a number of helpful definitions related to particle systems and particle dynamics.

First, we formalize the idea of extracting partial information from a particle system by only considering some subset of the quantities of each particle type.
\begin{definition}
    Let $S=(I,\tau,\nu,G)$ be a particle system with $\tau(i)=(I_{P,i},\kappa_{i})$ and $\varsigma:I\rightarrow\mathcal{P}(\mathbb{N})$ be a function such that for all $i\in I$ it holds that $\varsigma(i)\subseteq I_{P,i}$. Then $S'=(I,\tau',\nu,G)$ is called the \textbf{subsystem} of $S$ with respect to $\varsigma$ (written as $S'=\textup{subsystem}(S,\varsigma)$) if for all $i\in I$ it holds that $\tau'(i)=(\varsigma(i),\kappa_{i}|_{\varsigma(i)})$. In other words, $S'$ is the particle system obtained by taking only those quantities for each type $\tau(i)$ where the respective index of the quantity is in $\varsigma(i)$.
\end{definition}
\begin{definition}
    Let $s$ be a state of a particle system $S=(I,\tau,\nu,G)$ and $S'=\textup{subsystem}(S,\varsigma)$ be a subsystem of $S$. Then $s'=(\sigma',g)$ is called the \textbf{substate} of $s$ with respect to $\varsigma$ (written as $s'=\textup{substate}(s,\varsigma)$) if $s'$ is a state of $S'$ and for all $i\in I$ it holds that $\sigma'(i)=\textup{map}(\sigma(i),p\mapsto p|_{\varsigma(i)})$.
\end{definition}
Next, we formalize the idea of a subsystem being invariant under a particle-parallel particle dynamics.
\begin{definition}
    Let $S=(I,\tau,\nu,G)$ be a particle system and let $d:\mathcal{I}(S)\rightarrow\mathcal{I}(S)$ be a particle dynamics that is particle-parallel under $\Pi$. Furthermore, let $i\in I$ be a particle type index and $\tau(i)=(I_{P},\kappa)$ be the corresponding particle type. We then call $I_{\textup{immut}}\subseteq I_{P}$ an \textbf{immutable subset of the particle type} $\tau(i)$ under $d$ if for all $i_{P}\in I_{\textup{immut}}$ it holds that
    \begin{align}
        \begin{split}
            % \forall&(\sigma,g)\in\mathcal{I}(S):\forall p\in\sigma(i)\forall i_{P}\in I_{\textup{immut}}:\\
            % &p(i_{P})=((\Pi(i))(p,(\sigma,g)))(i_{P}).
            ((\Pi(i))(p,s))(i_{P}) = p(i_{P})
        \end{split}
    \end{align}
    In other words, there must be no state of $S$ where $\Pi(i)$ would mutate the state of any particle of type $\tau(i)$ such that a quantity with an index in $I_{\textup{immut}}$ is changed.
\end{definition}
\begin{definition}
    Let $S=(I,\tau,\nu,G)$ be a particle system and let $d:\mathcal{I}(S)\rightarrow\mathcal{I}(S)$ be a particle dynamics that is particle-parallel under $\Pi$. Furthermore, let $\upsilon:I\rightarrow\mathcal{P}(\mathbb{N})$ be a function. Then we call $\mathrm{subsystem}(S,\upsilon)$ an \textbf{immutable subsystem} of $S$ under $d$ if for all $i\in I$ it holds that $\upsilon(i)$ is an immutable subset of the particle type $\tau(i)$ under $d$.
\end{definition}
The notion of an immutable subsystem can be used to express the idea of extracting from a particle-system state only information that is invariant under a specific particle dynamics. This allows us to formulate a data-race-free subclass of the particle-parallel particle dynamics defined in Def.\ \ref{def:particle_parallel} in the case of in-place state manipulation. Similar to Def.\ \ref{def:parallelizable_general} we demand that each quantity of any particle in the system must either remain unchanged by the dynamics, or, if it is mutated, may only influence the dynamics of the particle it is associated with while being irrelevant to the dynamics of any other particle. In other words, the particle dynamics has \textit{exclusive mutability}.
\begin{definition}\label{def:exclusive_mutability}
    Let $S=(I,\tau,\nu,G)$ be a particle system and let $d:\mathcal{I}(S)\rightarrow\mathcal{I}(S)$ be a particle dynamics that is particle-parallel under $\Pi$. Also, let $\upsilon:I\rightarrow\mathcal{P}(\mathbb{N})$ be a function such that $S_\textup{immut}=\textup{subsystem}(S,\upsilon)$ is an immutable subsystem of $S$ under $d$. We then say that $d$ possesses \textbf{exclusive mutability} via the immutable subsystem $S_\textup{immut}$ and the restricted update function $\Pi_{\textup{immut}}:I\rightarrow\bigcup_{i\in I}(\mathcal{I}(\tau(i))\times\mathcal{I}(S_{\textup{immut}})\rightarrow\mathcal{I}(\tau(i)))$ if
    \begin{align}
        \begin{split}
            % \forall &(\sigma,g)\in \mathcal{I}(S): \forall i\in I:\forall p\in\sigma(i):\\
            (\Pi(i))(p,s)=(\Pi_{\textup{immut}}(i))(p,\textup{substate}(s,\upsilon)).
        \end{split}
    \end{align}
    In other words, for a dynamics $d$ that is particle parallel under $\Pi$ to possess exclusive mutability, there must be another function $\Pi_{\textup{immut}}$ that can reproduce the results of $\Pi$ while only depending on quantities of the respective particle and those quantities of other particles that are immutable under $d$.
\end{definition}
The reasoning why this implies data-race freedom is the same as in the case of the borrow rules of Rust. If a particle quantity is immutable under a particle dynamics, there is no writing access to it that would be required for a data race. Conversely, if the quantity is mutated, its visibility is restricted to a single thread of execution, namely that of the particle the quantity is associated with. This makes a data race, which requires at least two threads to access the same value, impossible.

While in Def.\ \ref{def:exclusive_mutability} we have reached the goal of defining a general class of particle dynamics that we can parallelize without the risk of data races, its usefulness still remains to be proven. To show that real-world particle dynamics can be expressed within this class, we will look at a number of special cases of particle dynamics with exclusive mutability that relate directly to real-world applications.

First, we consider the special case of a particle dynamics with exclusive mutability where each evolved particle state only depends on one particle state in the original system, i.e., the particle dynamics only processes information local to a single particle.
\begin{definition}
    Let $S=(I,\tau,\nu,G)$ be a particle system and let $d:\mathcal{I}(S)\rightarrow\mathcal{I}(S)$ be a particle dynamics that possesses exclusive mutability under $\Pi_\textup{immut}$. Then we call $d$ a \textbf{particle-local} dynamics if for all $ i\in I$ there is a function $f_i:\mathcal{I}(\tau(i)) \rightarrow \mathcal{I}(\tau(i))$ such that $\Pi_\textup{immut}(i)=((p,s)\mapsto f_{i}(p))$. In other words, we drop the dependency of $\Pi_{\textup{immut}}(i)$ on the immutable substate of the system for all particle types $\tau(i)$.
\end{definition}
This kind of dynamics implies that no information is shared between particles or, in a more physical interpretation, that the particles do not interact. These kinds of dynamics can therefore be found, e.g., in the simulation of ideal gases.

In many real-world particle dynamics, however, we find that particles do interact with each other. To limit the computational expense of simulating these dynamics, the interaction of each particle is typically restricted to a small subset of the particle system. This means that information between particles is not shared over the whole system, but only between small groups of particles. To define this formally, we first introduce the analog of a power set for states of particle systems.

\begin{definition}
    \label{def:particle_system_powerset}
    Let $S=(I,\tau,\nu,G)$ be a particle system and $s=(\sigma,g)$ be a state of $S$. We define the \textbf{power set $\mathcal{P}(s)$ of the particle-system state} $s$ as the set of all particle-system states\footnote{Note that in general $s'$ is not a state of $S$ in this context.} $s'=(\sigma',g)$ with $\sigma':I\rightarrow\mathcal{M}(\mathcal{I}(\mathfrak{P}))$ which fulfill the condition that for all $i \in I$ it holds that $\sigma'(i)$ is a submultiset of $\sigma(i)$. In physical terms, this means that all particles found in $s'$ appear in the same state in $s$, but not vice versa.
\end{definition}

Definitions \ref{def:particle_system_powerset} allows us to formalize the idea that calculating new particle states may not require knowledge of the entire immutable substate of the particle system, but only of a (potentially small) selection of particles from this state.

\begin{definition}
    Let $S=(I,\tau,\nu,G)$ be a particle system and let $d:\mathcal{I}(S)\rightarrow\mathcal{I}(S)$ be a particle dynamics that possesses exclusive mutability via the immutable subsystem $S_\textup{immut}$ and the restricted update function $\Pi_{\textup{immut}}$. Furthermore, let $\gamma_{ij}:\mathcal{I}(\tau(i))\times\mathcal{I}(\tau(j))\times G\rightarrow\mathbb{B}$ for all $i,j\in I$ be a family of Boolean predicates that encode group association between particles. Then we call $d$ a \textbf{group-local} dynamics under $\Pi_g$ with respect to $\gamma_{ij}$ if there is a function $\Pi_{g}:I\rightarrow\left(\bigcup_{i\in I}(\mathcal{I}(\tau(i))\times\{\mathcal{P}(s)|s\in\mathcal{I}(S_{\textup{immut}})\}\rightarrow\mathcal{I}(\tau(i)))\right)$ such that
    \begin{align}
        &(\Pi_{\textup{immut}}(i))(p,(\sigma_{\textup{immut}},g)) \nonumber\\
            &=(\Pi_{g}(i))(p,j\mapsto\textup{select}(\sigma_{\textup{immut}}(j),p'\mapsto\gamma_{ij}(p,p')),g).
    \end{align}
    In other words, we restrict the dependencies of $\Pi_{\textup{immut}}$ such that for each particle only information from other particles for which $\gamma_{ij}$ indicates group association is necessary to calculate the new state of the particle.
\end{definition}
An important detail in this definition is the fact that not just the evolved state, but also group association, must be decidable based on the information in an immutable substate of the particle system to prevent data races. It should also be noted that every particle dynamics that possesses exclusive mutability is a group-local dynamics with respect to \textit{some} group predicate, e.g., in the case of a particle-local dynamics a group predicate $\gamma_{ij}(p,p',g)=0$ or in the case of unrestricted particle interaction $\gamma_{ij}(p,p',g)=1$. However, the formalism allows us to express two other relevant subclasses of particle dynamics with exclusive mutability besides particle-local dynamics.

On the one hand, we can have group-local dynamics where the group association of particles is determined by some form of particle identity.
\begin{definition}
    Let $S=(I,\tau,\nu,G)$ be a particle system and let $d:\mathcal{I}(S)\rightarrow\mathcal{I}(S)$ be a particle dynamics that is group-local with respect to $\gamma_{ij}$. Furthermore, let $\textup{id}_{i}:\mathcal{I}(\tau(i))\rightarrow\mathbb{N}$ be a family of identifying function for all $i\in I$ such that
    \begin{align}
        \forall &i\in I:\forall k\in\mathbb{N}:\forall(\sigma,g)\in\mathcal{I}(S):\nonumber\\
        &|\textup{select}(\sigma(i),p\mapsto(\textup{id}_{i}(p)=k))|\leq1.
    \end{align}
    In other words, for every possible state of $S$ the function $\textup{id}_{i}$ must map every possible state of a particle of type $\tau(i)$ to a unique number.
    We then call $d$ a \textbf{fixed-group local} particle dynamics if there is a function $\gamma_g:\mathbb{N}^4\rightarrow\mathbb{B}$ such that $\gamma_{ij}(p,p',g)=\gamma_{\textup{group}}(i,\textup{id}_{i}(p),j,\textup{id}_{j}(p'))$. Conceptually, this means that the decision of group association does not need to inspect the whole state of each particle, but only the unique identifications produced by the function family $\textup{id}_{i}$ and the indices of the respective particle types.
\end{definition}
This class of particle dynamics is commonly used to enforce an association between particles that cannot be derived from within the simulation context and is therefore imposed on the system extrinsically. Typically these kinds of associations between particles are also considered unbreakable within the context of the particle simulation, which gives rise to the name of this category of particle dynamics. The most notable example for this are (fixed) chemical bonds. Notably, unlike many other molecular dynamics models, we make no assumption on the number of atoms involved in the bond. Typically, groups of up to four atoms (dihedral angles) are considered in molecular dynamics.

On the other hand, we can also have dynamic group associations based on parts of the particle state that vary over the course of a many-particle simulation. Since particle interactions typically decay with distance between particles, by far the most common case of this is a group predicate using the distance between particles.
\begin{definition}\label{def:neighborhood_local_dynamics}
    Let $S=(I,\tau,\nu,G)$ be a particle system and let $d:\mathcal{I}(S)\rightarrow\mathcal{I}(S)$ be a particle dynamics that is group-local with respect to $\gamma_{ij}$. Furthermore, let $r_\textup{cut}\in \mathbb{R}$ be a cutoff distance and $G \in \mathbb{B}^{\left\Vert I \right\Vert \times \left\Vert I \right\Vert}$ Boolean matrix indicating which particle types are allowed to interact. We then call $d$ a \textbf{neighborhood-local} dynamics if
    \begin{align}
        \gamma_{ij}(p,p',g)= G_{ij} \wedge \left\Vert \textup{pos}(p)-\textup{pos}(p')\right\Vert <r_{\mathrm{cut}}.
    \end{align}
\end{definition}

In practice, usually a special case of group-local dynamics is used in the form of \textit{pairwise} particle interactions, e.g., in the form of pairwise forces between particles. These interactions represent a particular method of calculating the evolved state of each particle from the state of all particles interacting with it.

First, we find each particle that is interacting with the particle being updated. Then, we calculate a single quantity for each of these interaction partners using only the initial state of the particle being updated and the interaction partner. Finally, we use some reduction scheme to calculate the evolved state of particle being updated from its initial state and all of the previously calculated pair quantities. In the example case of pairwise forces, this corresponds to first calculating the force for each pair of interacting particles and then summing these forces to obtain the net force for each particle. We can formalize this concept as follows.
\begin{definition}\label{def:pairwise_interaction}
    Let $S=(I,\tau,\nu,G)$ be a particle system and let $d:\mathcal{I}(S)\rightarrow\mathcal{I}(S)$ be a particle dynamics that is group-local under $\Pi_g$. Furthermore, for $i,j\in I$ let $Q_{ij}$ be a family of physical quantity types, $\mu_{ij}:\mathcal{I}(\tau(i))\times\mathcal{I}(\tau(j))\rightarrow Q_{ij}$ be a function family to map pairs of particle states to physical quantities and $\rho_{i}:\mathcal{I}(\tau(i))\times\left(I\rightarrow\bigcup_{j\in I}\mathcal{M}(Q_{ij})\right)\rightarrow\mathcal{I}(\tau(i))$ be family of reduction functions. We then call $d$ a \textbf{pairwise interaction} with the mapping functions $\mu_{ij}$ and the reduction functions $\rho_i$ if
    \begin{align}
        % \forall &i\in I:\forall\mathcal{I}(\tau(i)):\forall\sigma_{g}\in\{\mathcal{P}(s)|s\in\mathcal{I}(S_{\textup{immut}})\}: \nonumber\\
        (\Pi_{g}(i))(p,\sigma_{g})=\rho_{i}\left(p, j\mapsto\textup{map}(\sigma_{g}(j),p'\mapsto\mu_{ij}(p,p'))\right).
    \end{align}
\end{definition}

In practice, pairwise interactions are often also neighborhood-local, e.g., in the case of the popular Weeks-Chandler-Anderson potential for particle interactions.

When trying to map real-world particle simulations into the models developed in this section, we typically cannot find a suitably specialized classification of the \textit{entire} particle simulation for parallel execution. However, as stated before, real world particle simulations are often composed of multiple sub-dynamics which can be expressed as, e.g., particle-local dynamics or pairwise interactions. In these cases, one can use the parallel schemes developed here as long as there is a synchronization barrier between each sub-dynamics.\footnote{This is a result of the third condition for data races demanding that the memory accesses must be unsynchronized.}

\begin{figure}
    \includegraphics*[]{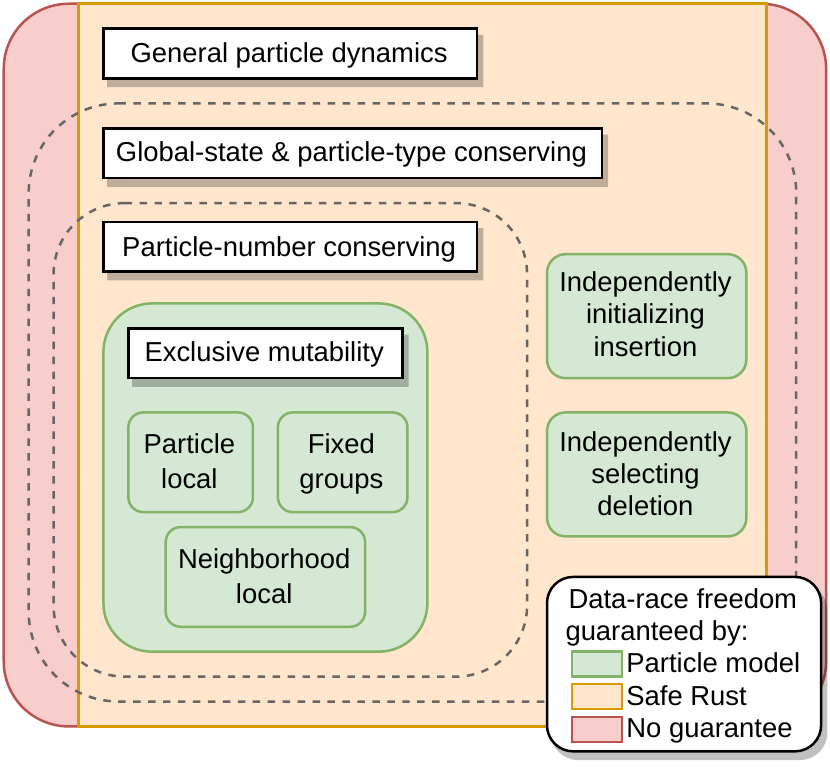}
    \caption{Overview of the different kinds of particle dynamics characterized in this article. The colors indicate how data-race freedom can be achieved in an implementation of this model, if at all possible. The green areas are inherently free from data races by their formulation in this article. The orange region indicates possible data-race freedom by choosing (Safe) Rust (or any language with equivalent safety guarantees) as an implementation language, while the red region indicates the possibility of particle dynamics for which neither our models nor the language model of Safe Rust can make statements regarding data-race freedom.}
    \label{fig:dynamics_levels}
\end{figure}

An overview of the different kinds of particle dynamics presented in this section and their relationship towards one another can be found in Fig.\ \ref{fig:dynamics_levels}.

\section{A domain-specific language for data-race-free particle simulations}\label{sec:DSL}
In this section, we present a practical application of the results of Sec.\ \ref{sec:data_race_free_simulations} in the form of a domain-specific language for many-particle simulations that can be parallelized while guaranteeing the absence of data races. First, we define the grammar of this language in Sec.\ \ref{subsec:grammar} before discussing how to ensure deadlock freedom in Sec.\ \ref{subsec:dead_lock_elimination} and finally presenting a number of example simulation codes to demonstrate the practicability of our approach in Sec.\ \ref{subsec:code_examples}.

\subsection{Language grammar}\label{subsec:grammar}
The purpose of this subsection is to translate the abstract models developed in Sec.\ \ref{sec:data_race_free_simulations} into the practical form of a (minimalistic) programming language for expressing concrete particle dynamics in a fashion that can then be compiled into a runnable program. Much like the concepts in Sec.\ \ref{sec:data_race_free_simulations} lean heavily on the borrow rules of Rust, here we lean on Rust also on a syntactical level. To describe this formally, we use a slightly modified version of the EBNF notation (see Appendix \ref{appendix_ebnf} for an overview) to define the grammar of the language. It should be noted that the language we describe in the following is meant to be stand-alone, but rather supposed to cooperate with some kind of environment hosting it. This greatly simplifies the design as we can delegate tasks requiring general computation interfaces to this hosting environment.

It is useful to define some common syntax elements first before getting to the more unique design points of the language. For one, we define the following basic non-terminals by regular expressions in the syntax of \textit{Perl compatible regular expressions} (PCRE) \cite{PCRE}
\begin{verbatim}
Identifier = [A-Za-z][A-Za-z0-9_]*
Nat = 0|([1-9][0-9]*)
Integer = (+|-)(0|([1-9][0-9]*))
Float = [+-]?[0-9]+(\.[0-9]*([eE][+-]?[0-9]+)?)?
\end{verbatim}
which in order of appearance denote identifiers (i.e., symbol names), natural numbers, integer numbers, and floating point numbers.

Another common syntax element is that of an \textit{expression}, which we define by the EBNF rule
\begin{verbatim}
Expression =
  Expression ("+" | "-" | "*" | "/") Expression |
  Identifier "(" (Expression ** ",") ")" |
  Identifier ("." Identifier)? ("[" Nat "]")? |
  "[" (Expression ** ",") "]" |
  Float | Integer
\end{verbatim}
where the productions (in order of appearance) encode binary arithmetic expressions, function calls, variables (including options for namespacing and static indexing), array literals, and primitive number literals. In terms of semantics, the only noteworthy aspect of expressions is that for arithmetic expressions involving array types both vector addition and scalar multiplication are defined. Otherwise the typical semantics for imperative languages apply.

Finally, we define types via the EBNF rule
\begin{verbatim}
Type = "i64" | "f64" | "[" Type ";" Nat "]" |
  "position"
\end{verbatim}
which in order of appearance correspond to 64-bit integer numbers, 64-bit floating point numbers, array types of fixed length, and the special \texttt{position} type which is equivalent to an array type of floating point numbers with a length corresponding to the dimensionality of the system. The \texttt{position} type might seem redundant, but compared to the general array type it encodes additional semantic meaning as it can be used to mark a quantity as the distinguished position quantity of a particle type.

From these syntax elements, we first construct a representation of the global-state space introduced in Def.\ \ref{def:particle_system}. Unless stated otherwise, the syntax constructs introduced in the following are all top-level constructs, i.e., can coexist simply separated by whitespace in a single unit of compilation. While for considerations of data-race freedom modeling the global state as a single lumped vector was sufficient, for practical purposes it is more sensible to allow splitting this vector into multiple distinct physical quantities. These can be expressed syntactically as a top-level construct similar to that of global variables:
\begin{verbatim}
GlobalStateMember = "global" Identifier
  ":" "mut"? Type ";"
\end{verbatim}
Here, the optional \texttt{mut} keyword indicates whether this quantity of the global state can be changed after compilation.\footnote{This allows certain optimization strategies such as constant folding or rewriting of expensive operations like divisions as a sequence of cheaper operations like additions and bitwise shifts.}

While the syntax of the global state we propose here is similar as for global variables, the semantics is not. As discussed in Sec.\ \ref{subsec:particle_dynamics}, the evolution of the global state can be better described by a general-purpose language. Therefore, we delegate the task of initializing and mutating the global state to the hosting environment.

Next, we look at the state space spanned by each particle type, which consists of an indexed set of physical quantities. For practical purposes, it is preferable to use symbolic names over numeric indices to distinguish both the quantities within a particle type and the particles types within a particle system. With this we can use a syntactical construct similar to heterogeneous product types (i.e., \texttt{struct} types) in Rust:
\begin{verbatim}
ParticleDefinition = "particle" Identifier
  "{" (ParticleQuantity ** ",") "}"
ParticleQuantity = Identifier ":" ("mut")? Type
\end{verbatim}
The optional \texttt{mut} qualifier is again a concession to possible optimizations such as eliding redundant storage of quantities that are constant for all particles of the same type. It should be noted that while the syntax is similar to \texttt{struct} types in Rust, the semantics is very different as the definition of a particle type neither introduces a new data type nor makes any statements about the layout of particle data in memory.

With the definitions of the static state structure being complete, we proceed with the syntactic encoding of particle dynamics.

As discussed at the end of Sec.\ \ref{subsec:particle_dynamics}, real-world particle dynamics typically consist of a loop containing multiple of the different classes of dynamics discussed in Sec.\ \ref{subsec:particle_dynamics}. To express this syntactically, a simulation \textit{schedule} is required that defines each of the loop iterations for each particle type. It is also useful to allow filtering of dynamics based on the loop counter (i.e., the simulation step number), e.g., to switch on and off certain parts of the dynamics at different times or to schedule evaluation steps that calculate statistical data from the state of the particle system. Overall, we propose the following syntax rules:
\begin{verbatim}
Simulation = "simulation" Identifier
  "{" ScheduleFilter* "}"
ScheduleFilter =
  "once" Nat "{" ParticleFilter "}" |
  "step" StepRange "{" ParticleFilter "}"
StepRange = Nat | Nat? ".." Nat? ("," Nat)?
ParticleFilter = "particle" Identifier
  "{" (Statement ** ";") "}"
\end{verbatim}
A simulation is thus described as a two-layered nested structure of \textit{statement} blocks where the first layer filters by simulation step and the second layer filters by particle type. For the simulation step filters, up to three numbers are required to express a step range where starting at step $a$ and ending at step $b$ (not inclusive) every $n$-th step is taken. We suggest that a single number is to be interpreted as a value of $n$ with $a=0$ and $b=\infty$, while the notation with up to three numbers should indicate $a$, $b$ and $n$ in this order with defaults of $a=0$, $b=\infty$, and $n=1$ being used if the respective number is omitted. The particle filter construct should express that the contents of its block of statements only applies to the particle type indicated by its symbolic name. To decide whether a simulation should terminate, control should (temporarily) be passed back to the hosting environment after each iteration of the simulation loop.

Nested within the simulation schedule, we use a set of statements to both express particle-local dynamics and provide a mechanism to ``escape'' the restrictions of the syntax we propose to the more expressive host environment. To this end, we define the following syntax rules for statements:
\begin{verbatim}
Statement =
  "let" Identifier ":" Type "=" Expression |
  Identifier ("[" Nat "]")? "=" Expression |
  "call" Identifier |
  "update" Identifier ("." Identifier)?
\end{verbatim}
The first two rules for statements are used to implement particle-local dynamics with the first one allowing to introduce new (temporary) variables and the second one allowing to either write a value to one of the previously declared variables or to mutate a particle quantity of the particle type of the current particle filter. The optional natural number in square brackets can be used to assign values to individual fields of vector variables or quantities.

Statements beginning with the keyword \texttt{call} indicate that control should be handed over to the hosting environment temporarily at this point in the simulation. This is likely to be implemented as a form of \textit{callback}, i.e., a function pointer from the hosting environment registered under a symbolic name which can then be referenced in this kind of statement after the \texttt{call} keyword.

Finally, statements beginning with the \texttt{update} keyword implement general group-local interactions. Since interactions can be defined between two or more particle types, it is sensible to place these syntax constructs outside the simulation schedule (which is filtered by particle types) and reference them by a symbolic name after the \texttt{update} keyword. In this article, we only develop syntax for neighborhood-local, pairwise interactions and fixed-group local dynamics with a constant group size known at compile time as these are by far the most common primitives for molecular dynamics in particular and can be implemented with a relatively simple grammar. Both of these can be represented as follows:
\begin{verbatim}
PairInteraction =
  "pair interaction" Identifier "("
    Identifier ":" Identifier ","
    Identifier ":" Identifier
  ")" "for" ("|" Identifier "|" "=")?
    Identifier "<" Expression
  InteractionBody
FixedGroupInteraction =
  "fixed group interaction" Identifier "("
    (Identifier ":" Identifier) ** ","
  ")"
  InteractionBody
\end{verbatim}
Here, the dynamics is first bound to a name following the \texttt{pair interaction} or \texttt{fixed-group interaction} keywords, respectively. After this, the two interacting particle types for pairwise interactions or an arbitrary number of particle types are specified. This is done with a syntax similar to function parameters in Rust as a pair of identifiers per particle type, where the first identifier defines a namespace for the quantities of each of the types and the second identifier specifies the particle type by its symbolic name. This again represents a syntactic similarity of particle types and structured data types even though the semantics of both are very different. Binding a namespace to each particle involved in the dynamics is necessary as there is no guarantee that quantity names are unique between different particle types.

In the case of a pairwise interaction, following the particle type specification, after the \texttt{for} keyword is the required information for defining the particle neighborhood, which, as discussed in Def.\ \ref{def:neighborhood_local_dynamics}, is defined by a cutoff distance. This distance is given as an expression after the symbol \texttt{<}. An implementation has to verify that this expression evaluates to a numeric value and only depends on constant quantities from the global state. Before this, one or two identifiers have to be placed in a notation resembling the equation $\left\Vert \vec{r}\right\Vert =r<r_{\text{cut}}$. Both of these identifiers serve the purpose of binding symbolic names to the distance between two interacting particles with the first, optional identifier denoting the distance \textit{vector} and the second identifier denoting the \textit{scalar} distance. Binding the distance (vector) to a name in such a prominent position is motivated by the fact that many-particle interactions depend on the inter-particle distance to determine the interaction strength. In the case of periodic boundary conditions, this also allows to differentiate between the position of the particle in the simulation domain and the distance vector that might correspond to one of the images of the particle.

The last part of both the definition of a pairwise interaction and a fixed-group local dynamics is a section containing the information on what calculations are involved in the respective dynamics and which quantities are to be mutated. We propose the following syntax rules for this task:
\begin{verbatim}
InteractionBody = "{"
  ("common" "{"
  (InteractionStatement ** ";")*
  "}")?
  InteractionQuantity*
"}"
InteractionQuantity =
  "quantity" Identifier
  "-[" ReductionMethod "]->" TargetSpecifier "{"
    (InteractionStatement ** ";")+ ";"
    Expression
  "}"
InteractionStatement =
  "let" Identifier ":" Type "=" Expression |
  Identifier ("[" Nat "]")? "=" Expression
ReductionMethod = "sum" | "max" | "min"
TargetSpecifier =
  (("-")? Identifier "." Identifier)
    ** ("=" | ",")
\end{verbatim}
Here, an \texttt{InteractionBody} is composed of one optional \texttt{common} block as well as multiple definitions of \textit{interaction quantities}, i.e., the physical quantities calculated for every pair of interacting particles. The purpose of the \texttt{common} block is to allow the definition of variables used for the calculation of more than one interaction quantities. The definitions of interaction quantities follows very closely to Def.\ \ref{def:pairwise_interaction}.

Following the symbolic name of the interaction quantity is a specification of the reduction method, i.e., the equivalent to the function $\rho_i$ in Def. \ref{def:pairwise_interaction} which transforms the individual results of all pairwise interactions of a particle into a single physical quantity. Here, we suggest only three common reduction operations as examples, namely summing all pairwise interaction results or taking either the minimum or maximum of all interaction results.

After the specification of the reduction method follows a specification of the target particle quantities that this interaction quantity should be written to. The syntax for the target quantities can be seen as a variation of the pattern-matching syntax of Rust. Here, this pattern is matched against the result of the last expression in the block following the target specification. The target specification itself is composed of multiple namespaced particle quantities separated either by equality signs or commas. Here, an equality sign indicates that a single result from the calculation of the interaction quantity is written to multiple particle quantities, while a comma indicates that the result is an array which is to be split into its elements which are then written to distinct particle quantities. Both variants have the option of negating the result, which is a convenience for force-like quantities which are subject to Newton's third law as these can then simply be expressed as \texttt{p1.F = -p2.F} for two particle namespaces \texttt{p1} and \texttt{p2} each of which contains a particle quantity \texttt{F}. Notably, target specifiers can contain both target quantities separated by equality signs and targets separated by commas.\footnote{This is only useful for fixed-group local dynamics as pairwise interactions cannot have more than two targets.} For example, an interaction quantity for an interaction of three particles with namespaces \texttt{p1}, \texttt{p2}, and \texttt{p3}, respectively, and a particle quantity \texttt{F} in each namespace can have the target specifier \texttt{p1.F = -p3.F, p2.F}. This indicates that the calculation of the interaction quantity produces a vector of two values, the first of which is written to \texttt{p1.F} and in negated form to \texttt{p3.F} while the second one is written to \texttt{p2.F}.

Finally, after the target specifier, the actual calculation of each interaction quantity follows in the form of a block of statements terminated by an expression that determines the value of the interaction quantity. In this context, the statements are limited to variable definitions and assignments. In all expressions appearing in this section, it is essential to respect the constraints of Def.\ \ref{def:exclusive_mutability}, i.e., to forbid access to any particle quantities that are mutated by this interaction, i.e., all quantities appearing as part of a target specification in this interaction.

It should be noted that the fact that each interaction quantity can only be written to a single particle quantity is a restriction compared to the more general reduction function $\rho$ in Def.\ \ref{def:pairwise_interaction}. The reason for this is the necessity for an implementation to statically analyze which quantities of each particle is immutable under the interaction and therefore visible in the computation of the interaction quantities. With the grammar above this task becomes trivial as the programmer must explicitly indicate that a variable is not immutable. A similar mandatory opt-in for mutability can also be found in the design of Rust.

For the two data-race-free, particle number altering dynamics -- namely independently initializing insertions and independently selecting deletions -- it is necessary to implement them in particularly tight integration to the hosting environment as they often require general purpose functionality (such as I/O operations to load data for initialization or random number generation for probabilistic deletion of particles). Therefore, we do not propose a special syntax for these dynamics but assume that they can be implemented by a suitable mechanism in the hosting environment (e.g., function pointers for dynamic dispatch or generics for static dispatch).

One aspect a reader might find surprising in the syntax developed in this section is the lack of any control-flow structures such as conditionals or loops. The primary reason for this is the fact that branching control flow makes static program analysis much harder if not impossible. In particular, for reasons explained in the following section, it is required that the sequence of elementary particle dynamics for each particle type must be known at compile time to guarantee the absence of deadlocks. Allowing conditional control-flow would require verifying all possible control flow paths which is impractical as their number grows exponentially with the number of control flow statements. This problem could be worked around by preventing conditional control flow to ``leak'' beyond the boundaries of each elementary particle dynamics, e.g. by using syntax elements such as ternary operators to allow branching control flow only on the level of expressions. However, even in this case branching code can still be problematic as it interferes with optimization strategies such as vectorization and adds complexity to the optimization tasks due to effects of branch prediction. Therefore, for very complex branching code it might be more viable to delegate this part of the dynamics to the hosting environment instead.

\subsection{Eliminating interaction deadlocks}\label{subsec:dead_lock_elimination}

As discussed at the end of Sec.\ \ref{subsec:particle_dynamics} for particle dynamics composed of multiple (sub-)dynamics, it is necessary to add synchronization barriers between each sub-dynamics for each particle type. As interactions can affect more than one particle type, this, however, also synchronizes the simulation schedules of multiple particle types at certain points. So far, we implicitly assumed that an implementation will ensure that these synchronization points match up, i.e., that if an interaction between a particle type \texttt{A} and a particle type \texttt{B} is implied in the simulation schedule of type \texttt{A} there is a corresponding one found in the schedule for type \texttt{B}. This, however, brings with it the problem of \textit{deadlocks}, i.e., multiple interactions blocking each other from progressing in a dependency cycle. For example, consider the following simulation:
\begin{verbatim}
particle A { /* ... */ }
particle B { /* ... */ }
particle C { /* ... */ }

pair interaction AB(a: A, b: B) {}
pair interaction BC(b: B, c: C) {}
pair interaction CA(c: C, a: A) {}

simulation {
  step {
    particle A {
      update AB;
      update CA;
    }
    particle B {
      update BC;
      update AB;
    }
    particle C {
      update CA;
      update BC;
    }
  }
}
\end{verbatim}
This simulation will stall indefinitely as each simulation schedule prescribes a different interaction to be performed first.

\begin{figure}
    \includegraphics[]{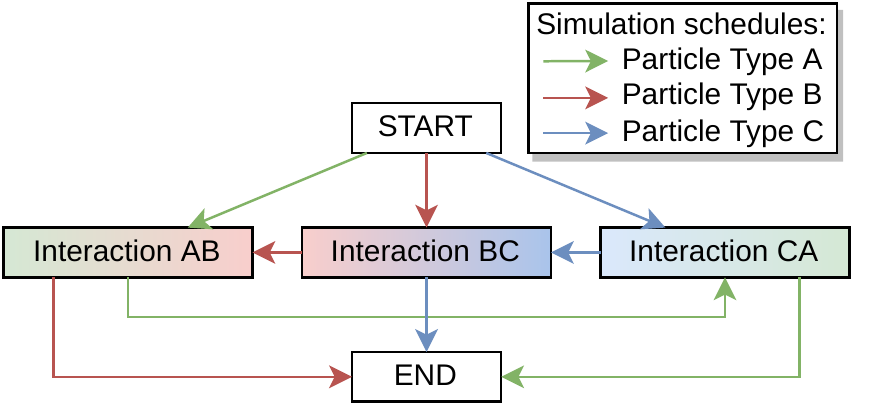}
    \caption{Example of an interaction graph of a particle simulation of three particle types. The color of the arrows indicates the particle type to which the simulation schedule for this directed edge belongs. The clearly visible cycle in the directed graph means that this simulation contains a deadlock and is therefore invalid.}
    \label{fig:deadlock}
\end{figure}

To detect the presence of deadlocks in a simulation, we can use a graph-based intermediate representation of the simulation schedule. The nodes of this graph represent each interaction in the simulation schedule as well as by two special nodes the start and end of each simulation step. Then, for each particle type a directed path is added beginning at the start node and connecting the interaction nodes in the order they appear in the simulation schedule for this particle type. Each of these paths is then terminated at the end node. Such an \textit{interaction graph} is presented in Fig. \ref{fig:deadlock} for the previous example.

Since interaction deadlocks are created by a cyclic dependency of interactions on one another, a simulation is deadlock-free if the \textit{interaction graph} is acyclic. This can be verified by standard methods such as Tarjan's algorithm \cite{Tarjan1972}.

\subsection{Examples}\label{subsec:code_examples}

In this section, we provide examples of common simulations to illustrate the application of the previous results to real-world particle dynamics.

\subsubsection*{Newtonian dynamics}
First, we present in Listing \ref{code:velocityverlet} an example for a particle system subject to Newtonian dynamics. We employ the commonly used velocity-Verlet scheme to solve these dynamics numerically. The particle system itself is composed of free moving particles using the Weeks-Chandler-Andersen (WCA) potential \cite{Weeks1971}
\begin{equation} \label{eq:wca}
U(r)=\begin{cases}
    4\varepsilon \left[ \left( \frac{\sigma}{r} \right)^{12} - \left( \frac{\sigma}{r} \right)^{6} \right] + \varepsilon & \mbox{if } r \leq 2^{1/6} \sigma, \\
    0 & \mbox{else}
    \end{cases}
\end{equation}
to model particle interactions. Here, $r$ is the particle distance, $\varepsilon$ is an energy scaling factor, and $\sigma$ is the diameter of the particles.

In the implementation, we first make use of the global state to define the time-step size, the particle diameter $\sigma$, and the energy scaling factor $\varepsilon$ of the WCA potential as constants of the simulation. This allows us to insert the concrete values for these constants from the hosting environment without having to change the domain-specific simulation code.

Next, we define a particle type representing a spherical particle without orientational degrees of freedom. The physical state of the particle is fully described by its position, velocity, and mass, the latter of which is considered to be constant. Additionally, we define particle quantities for the net force acting on the particle and the total potential energy associated with the particle. These two quantities are not necessary to describe the physical state of the particle, but rather store the physical quantities resulting from particle interactions.

Afterwards, we define a pairwise particle interaction between two particles of the previously defined particle type with the cutoff distance defined by the WCA potential. After defining a variable for the common term $(\sigma/r)^6$, we define two interaction quantities. The first quantity is the repulsion force defined by $-\vec{\nabla}U$, which is written to the force quantity of each particle with a negation for one of the particles to account for Newton's third law. The second quantity calculates the potential energy of the interaction evenly split between both particles. This allows to calculate the total potential energy of the system by summing over all particles.

Finally, we define a simulation schedule for the \texttt{Spherical} particle type in combination with the velocity-Verlet integration scheme. Here, the particle dynamics is composed of a particle-local dynamics mutating position and velocity, a pairwise interaction that mutates forces and potential energy, and then another particle-local dynamics mutating velocity.

Overall, this example shows that even though the domain-specific language has no in-built support for any concrete numerical scheme, it can still express real-world simulations relatively concisely. For example, the entire velocity-Verlet scheme only takes up four lines of code.

\begin{lstlisting}[float=*,label={code:velocityverlet},
    caption={Example of a Newtonian dynamics simulation of a system of spherical, monodisperse particles subject to a repulsive force based on the Weeks-Chandler-Andersen potential in three spatial dimensions. Here, the velocity-Verlet scheme is used to integrate the equations of motion. The \texttt{pow} function used here is defined as the result of taking the first argument to the power of the second argument. In addition to each particle trajectory, this simulation also calculates the potential energy of each particle by splitting the potential energy of each particle interaction evenly between particles.}
,basicstyle=\ttfamily]
global DT: f64;      // Velocity-Verlet time step
global SIGMA: f64;   // Particle diameter
global EPSILON: f64; // WCA scaling factor

// Representation of spherical particles
particle Spherical {
    x : mut position,
    v : mut [f64; 3], // Velocity of particle
    F : mut [f64; 3], // Total force acting on particle
    U : mut f64,      // Total potential energy of particle
    mass: f64         // Particle mass
}

// Pairwise repulsion of particles via the WCA potential
pair interaction Repulsion (p1: Spherical, p2: Spherical)
    for |rvec| = r < pow(2, 1.0 / 6.0) * SIGMA
{
    common {
        // Inverse of inter-particle distance relative to particle diameter
        let relative_distance_inverse = pow(SIGMA/r,6);
    }

    // Force derived from WCA potential
    quantity force_WCA -[sum]-> p1.F = -p2.F {
        let scale: f64 = -4.0 * EPSILON * (
            12.0 * relative_distance_inverse * relative_distance_inverse
            - 6.0 * relative_distance_inverse) / (r * r);
        scale * rvec
    }

    // Potential energy from WCA potential (split evenly between particles)
    quantity potential_WCA -[sum]-> p1.U = p2.U {
        0.5 * (4.0 * EPSILON * (
            relative_distance_inverse * relative_distance_inverse
            - relative_distance_inverse
        ) + EPSILON)
    }
}

// Velocity-Verlet-based simulation of Newtonian particle dynamics
simulation VelocityVerlet {
    step 1 {
        particle Spherical {
            v = v + 0.5 * F / mass * DT;
            x = x + v * DT;
            update Repulsion;
            v = v + 0.5 * F / mass * DT;
        }
    }
}
\end{lstlisting}

\subsubsection*{Harmonic valence angle potential}

Next, we look at the example in Listing \ref{code:angularbond} which presents an implementation of a chemical bond between three atoms with a harmonic potential for the valence angle, i.e.,
\begin{equation}
    U(\theta) = k(\theta - \theta_0)^2,
\end{equation}
where $\theta$ is the valence angle, $k$ is the bond strength, and $\theta_0$ is the equilibrium valence angle. This is an important example as potentials depending on bond angles are a very common archetype in molecular dynamics.

In this example, we reuse the same particle type as in Listing \ref{code:velocityverlet} for spherical particles and use a similar style of passing the constants $k$ and $\theta_0$ to the simulation via global state quantities. After this we define a fixed-group interaction for triples of particles. This interaction defines a single interaction quantity that writes three different forces to each of the three particles involved in the interaction. For the sake of clarity, we use the formulation of Ref.\ \cite{Monasse2014} for calculating the forces on each atom. This is algebraically equivalent, but numerically less efficient than the typical formulation found in molecular dynamics software packages for (harmonic) angular bonds.

Particularly noteworthy is the \texttt{distvec} function in this example. Typically, one would calculate the distance vector between two particles by a simple vector difference of the position vectors of each particle. However, this method is problematic if periodic boundary conditions are used, in which case one has to decide the image of each particle that is used for the distance calculation. For this purpose, we propose the implementation to support the pseudo function \texttt{distvec} that takes two particle namespaces and returns the distance vector for the particles they represent. In the case of periodic boundary conditions, we can define the distance vector between particles as the vector between the position of the first particle and the position of the closest image of the second particle (\textit{minimum image convention} \cite{Deiters2013}).

Overall, this example shows one of the more common applications of fixed-group interactions in the form of a potential depending on bond angles. It can be extended easily to larger particle groups, e.g., to implement potentials depending on dihedral angles.

\begin{lstlisting}[float=*,label={code:angularbond},
    caption={Definitions of a particle type for spherical particles and a three-particle interaction emulating a chemical bond with a harmonic potential for the valence angle in three spatial dimensions. Typically, this interaction would be combined with an interaction related to the bond \textit{length}, which is omitted here for the sake of brevity. In the calculation of the interaction quantity \texttt{force\_harmonic}, we use the \texttt{length} function to calculate the length of a vector, the \texttt{acos} function to calculate the arccosine of its argument, and the \texttt{dot} and \texttt{cross} functions to calculate the dot and cross products of two vectors, respectively. We also use a special notation to determine the distance vector between two particles by means of a pseudo function \texttt{distvec}, which we assume the implementation will substitute with an appropriate calculation given the boundary conditions of the system. The calculation itself closely follows Ref.\ \cite{Monasse2014}.}
,basicstyle=\ttfamily]
global K: f64;       // Bond strength
global THETA_0: f64; // Bond equilibrium angle

particle Spherical {
    x : mut position,
    v : mut [f64; 3],
    F : mut [f64; 3],
    mass: f64
}

fixed-group interaction HarmonicAngle (p1: Spherical, p2: Spherical, p3: Spherical) {
    quantity force_harmonic -[sum]-> p1.F, p2.F, p3.F {
        // Distance vectors
        let r_21: [f64; 3] = distvec(p2,p1);
        let r_23: [f64; 3] = distvec(p2,p3);
        let d_21: f64 = length(r_21);
        let d_23: f64 = length(r_23);
        // Bond angle
        let theta: f64 = acos(dot(r_21,r_23) / (d_21 * d_23));
        // Directions of forces on p1 and p2
        let perp_1: [f64; 3] = norm(cross( r_21, cross(r_21,r_23)));
        let perp_3: [f64; 3] = norm(cross(-r_23, cross(r_21,r_23)));
        // Forces
        let f_1: [f64; 3] = -2.0 * K * (theta - THETA_0) / d_21 * perp_1;
        let f_3: [f64; 3] = -2.0 * K * (theta - THETA_0) / d_23 * perp_3;
        let f_2: [f64; 3] = -f_1 - f_2;
        [f_1, f_2, f_3]
    }
}
\end{lstlisting}

\subsubsection*{Brownian dynamics}

The final example is shown in Listing \ref{code:eulermaruyama}, which demonstrates how a simple Brownian dynamics can be implemented in form of the Euler-Maruyama scheme. Assuming spherical particles, the numerical scheme then reads
\begin{equation}\label{eq:brownian}
    \vec{r}_i(t+\Delta t) = \vec{r}_i(t) + \frac{\Delta t}{\gamma} \vec{F}_i (\vec{r}_i (t), t) + \sqrt{\frac{2k_\mathrm{B}T\Delta t}{\gamma}} \vec{\eta}_i,
\end{equation}
where $\vec{r}_i(t)$ is the position of the $i$-th particle at time $t$, $\Delta t$ is the time-step size, $\gamma$ is the friction coefficient, $\vec{F}_i (\vec{r}_i (t))$ is the total force acting on the $i$-th particle at time $t$, $k_\mathrm{B}$ is the Boltzmann constant, $T$ is the system temperature, and $\vec{\eta}_i$ is a Gaussian white noise vector with zero mean and a variance of one.

Again, the implementation begins with a declaration of the global state quantities, which in this case consist of $\Delta t$, $\gamma$, and $T$. Additionally, we use the global-state quantities to pass $k_\mathrm{B}$ as a symbolic value for the sake of brevity in this example. The particle type is notably shorter compared to the earlier examples since particle mass and instantaneous velocity are irrelevant in the case of Brownian dynamics\footnote{More precisely, inertial forces are considered negligible compared to friction forces.}. Finally, in the definition of the simulation schedule, we find an almost identical expression to Eq.\ (\ref{eq:brownian}).

This example demonstrates how the model of particle dynamics we propose can be molded to the exact requirements of a numerical scheme. In particular, one can define particle types with just the degrees of freedom a particle has under a given numerical scheme thus saving resources. For example, in Brownian dynamics there is no need to store the velocity of each particle. Of course, this places some responsibility on the programmer to implement a \textit{numerically} sound particle dynamics as the domain-specific language only prevents the user from formulating particle simulations that are unsound in the sense that they exhibit undefined behavior (e.g., data races).

\begin{lstlisting}[float=*,label={code:eulermaruyama},
    caption={Example of a Brownian dynamics simulation of a system of spherical particles. The exact definition of the forces acting on the particle are omitted for the sake of brevity. This example makes use of the \texttt{random\_normal} function which produces a single sample from a Gaussian white noise of mean zero and variance one as well as the \texttt{sqrt} function which calculates the square root of its argument.}
,basicstyle=\ttfamily]
global DT: f64;    // Euler-Maruyama time step
global GAMMA: f64; // Friction coefficient
global TEMP: f64;  // System temperature
global K_B: f64;   // Boltzmann constant

particle Spherical {
  x : mut position,
  F : mut [f64; 3]
}

simulation EulerMaruyama {
  step 1 {
    particle Spherical {
      // Interaction updates and external forces omitted for brevity
      // [...]
      let normal_noise: [f64; 3] = [random_normal(), random_normal(), random_normal()];
      x = x + DT / GAMMA * F + sqrt(2.0 * K_B * TEMP * DT / GAMMA) * normal_noise;
    }
  }
}
\end{lstlisting}

\section{Conclusions}\label{sec:conclusions}
We have developed a formalization of the concepts of generic particle systems and based on this a very general definition of particle dynamics. This was done in the context of safe parallelism, i.e., the avoidance of data races, to find a safe abstraction for computational implementations of particle dynamics in the shared-memory model of parallel programming. For this we took inspiration from the design of the general purpose programming language Rust which is able to guarantee the absence of data races by enforcing a set of borrowing rules at compile time. We found that general particle dynamics are not algorithmically constrained enough to transfer these rules directly. Instead, we formulated a hierarchy of restrictions on particle dynamics that eventually allowed us to define concrete subclasses of particle dynamics which we can safely parallelize automatically. In particular, we identified two classes of particle dynamics that alter the number of particles in a particle system in such a way that they can be split into independent per-particle operations. For particle dynamics that do not add or remove particles, we found an equivalent of the borrowing rules of Rust in the context of particle systems. This concept was then concretized into multiple subclasses of particle dynamics that are common in practical applications.

After this, we designed a domain-specific programming language around our abstract formulation and classification of particle dynamics. Borrowing heavily from the syntax of existing languages, we formulated a minimalist grammar for this language and discussed where static analysis is required within this language to ensure data-race freedom. We found that unlike Rust, which utilizes an elaborate type system to ensure safety, for the language we designed it is sufficient to regulate symbol visibility and perform basic type checking. We also showed that by means of an appropriate intermediate representation we can not only guarantee data-race freedom, but also deadlock freedom. To prevent over-extending our language and still overcome the inherently limited expressiveness of a domain-specific language, we designed the language to be embedded in a general-purpose programming language and added mechanisms for these two contexts to cooperate.

Finally, we asserted the practicability of our model by expressing common operations from molecular dynamics in our domain-specific programming language. These examples also demonstrate the high degree of flexibility our model provides which we expect to be very useful for prototyping more exotic many-particle dynamics.

There are many opportunities for future extensions of our results presented in this work. For one, we have only formulated safe abstractions for particle dynamics that are local either to a single particle or to a small group of particles. There are, however, many physical systems that include long-range interactions (e.g., long-range electrostatic interactions) which often require more sophisticated algorithms such as Ewald summation \cite{Ewald1921}, the particle-particle-particle-mesh method \cite{Hockney1988}, or the Barnes-Hut method \cite{Barnes1986} for simulations. More research is required to find general, data-race-free abstractions for these kinds of methods.

Naturally \textit{implementing} the language we designed here is another big task. A partial implementation of the model we designed here for multicore CPU-based systems can be found in Ref.\ \cite{FIPS}. Due to the high degree of parallelism we provide in our model of particle dynamics, the development of an implementation based on GPU hardware is another worthwhile avenue to continue this work.

There is also much opportunity for extending the syntax we gave in this article for real-world applications. Here, we presented a very minimal language model to ease the discussion, but this leaves much to be desired in terms of language features. Potential areas of improvement include extensions of the type system to support tensor quantities of a rank higher than one, support for sharing behavior between particle types to reduce redundant code, a module system to provide common primitives such as interactions for commonly used potentials, limited forms of conditional expressions and loops, and many more. Due to the already present possibility of integrating a general-purpose programming language into the model, strictly speaking these extensions do not add functionality, but will drastically increase the usability of any implementation of our model.

We expect that the model of particle dynamics and the language we developed for it will aid future research related to the numerical treatment of many-particle systems, and we hope that in return feedback from these applications will guide future refinement of abstractions for safe particle dynamics.

\begin{acknowledgments}
We thank Cornelia Denz, Matthias R\"uschenbaum, Stephan Br\"oker and Tobias Nitschke for helpful discussions. This work is funded by the DFG -- Project-ID 433682494 - SFB 1459.
\end{acknowledgments}

\appendix
\section{\label{appendix_multisets}Multiset notation}

This article makes extensive use of multisets for which there is no generally accepted notation. In this section, we therefore formally define multisets, their notation, and all operations on them used in this work.

\begin{definition}
    A \textbf{multiset} $M$ over a set $U$ is defined as a tuple $(U,\mu)$ composed of the \textbf{underlying set} $U$ and a \textbf{multiplicity function} $\mu:U\rightarrow\mathbb{N}_0$. $x\in U$ is an \textbf{element of} $M$ of multiplicity $n$ (written as $x\in^n M$ or $x\in M$ if the multiplicity is irrelevant) if and only if $\mu(x) = n > 0$.
    We represent a multiset by an enumeration of its elements, each repeated according to its multiplicity, in square brackets, e.g., the multiset $[1,1,1,2]$ over the underlying set $\{1,2\}$ is equivalent to $(\{1,2\},((1,3),(2,1)))$. $[\;]$ is the \textbf{empty multiset} which does not contain any elements.
\end{definition}

\begin{definition}
    We define $\mathcal{M}(U)$ as the set of all multisets over $U$.
\end{definition}

\begin{definition}
    Let $M$ and $M'$ be multisets. $M$ and $M'$ are equal (written as $M=M'$) if $x\in^n M\Leftrightarrow x\in^n M'$.
\end{definition}

\begin{definition}
    The \textbf{cardinality} of a multiset $M=(U,\mu)$ is defined as $|M|=\sum_{x\in U}\mu(x)$.
\end{definition}

\begin{definition}
    The \textbf{sum of two multisets} $M_1=(U,\mu_1)$ and $M_2=(U,\mu_2)$ over a common underlying set $S$ is defined as the multiset $M_1 + M_2 = (U,\mu_s)$ with $\mu_s:U\rightarrow \mathbb{N}_0, x\mapsto \mu_1(x) + \mu_2(x)$.
\end{definition}

\begin{definition}
    Let $M=(U,\mu)$ be a multiset. Another multiset $M'=(U,\mu')$ is called a \textbf{submultiset} of $M$ (written as $M'\subseteq M$) iff $\forall x\in U: \mu'(x) \leq \mu(x)$. We define the \textbf{powerset} $\mathcal{P}(M)$ as the sets of all submultisets of $M$, i.e., $M'\in\mathcal{P}(M)\Leftrightarrow M'\subseteq M$.
\end{definition}

\begin{definition}
    Let $M=(U,\mu)$ be a multiset over a set $U$ and $f:U\rightarrow U'$ a function. We define the \textbf{map} operation as $\textup{map}(M,f)=\sum_{x\in^{n}M}\sum_{j=1}^{n}[f(x)]$. This operation creates a new multiset over the set $U'$ from the result of mapping each element of $M$ to another element via the function $f$ while preserving multiplicity.
\end{definition}

\begin{definition}
    Let $M=(U,\mu)$ be a multiset over a set $U$ and $p:U\rightarrow \mathbb{B}$ be a Boolean predicate function.
    We then define the \textbf{select} operation as
    \begin{align}
        \textup{select}(M,p)=\sum_{x\in^{n}M}\sum_{j=1}^{n}\left( x \mapsto \begin{cases}
            [x]   &\textup{if } x\in M \wedge p(x)\\
            [\;]  &\textup{else}
        \end{cases} \right).
    \end{align}
    This operation determines the largest submultiset of $M$ which only contains elements fulfilling the predicate $p$.
\end{definition}

\section{\label{appendix_ebnf}Modified EBNF notation}

In this article we use a slight modification of the classical Extended Backus-Naur form (EBNF) syntax \cite{EBNF} to express the productions of a context-free grammar. In particular, we do not use the comma as a concatenation operator and instead simply use whitespace to separate the elements of a sequence. We also omit the terminating semicolon and indicate the end of a rule by indentation instead. Furthermore we do not use brackets and curly parentheses to denote repetition and instead use a similar notation to regular expressions, i.e., a question mark indicates that the preceding terminal or non-terminal is optional, a star indicates that it can be repeated an arbitrary time including zero, and a plus sign indicates arbitrary repetition with at least one occurrence. For the sake of brevity, we also introduce the double-star operator \texttt{**} which indicates that the previous terminal or non-terminal can be repeated arbitrary many times but each repetition is separated by the following terminal or non-terminal. As an example, the expression \texttt{"A" ** ","} can produce the empty word, the word \texttt{A}, the word \texttt{A,A}, the word \texttt{A,A,A}, etc.. Finally, we omit explicit whitespace terminals as they contribute little to the actual language design. In an implementation of the grammar one must therefore take care to introduce mandatory whitespace as is necessary to eliminate ambiguity, e.g., in places where an identifier can collide with a keyword.

\bibliography{refs}

\end{document}